# Ultra-High Mechanical Flexibility of 2D Silicon Telluride

Romakanta Bhattarai and Xiao Shen[*]

Department of Physics and Materials Science, University of Memphis, Memphis, Tennessee 38152, USA

**Abstract**

Silicon telluride ($Si_2Te_3$) is a two-dimensional material with a unique variable structure where the silicon atoms form Si-Si dimers to fill the "metal" sites between the Te layers. The Si-Si dimers have four possible orientations: three in-plane and one out-of-the plane directions. The structural variability of $Si_2Te_3$ allows unusual properties especially the mechanical properties. Using results from first-principles calculations, we show that the $Si_2Te_3$ monolayer can sustain a uniaxial tensile strain up to 38%, highest among all two-dimensional materials reported. The high mechanical flexibility allows applying mechanical strain to reduce the band gap by 1.4 eV. With increasing strain, the band gap undergoes an unusual indirect-direct-indirect-direct transition. We also show that the uniaxial strain can effectively control the Si-Si dimer alignment, which is beneficial for practical applications.

The research of two-dimensional materials has significantly advanced in the past decade. The examples include graphene,[1-3] transition-metal dichalcogenides such as $MoS_2$,[4] black phosphorous,[5,6] and atomic layers of topological insulators such as $Bi_2Se_3$.[7,8] With the reduced dimensionality, these materials allow a broad range of new opportunities not present in 3D systems, such as minimal dielectric screening,[9] rippling,[10] twisting,[11,12] and easy functionalization by chemical treatments[13,14] and irradiation.[15,16] These opportunities make the 2D materials particularly attractive for both fundamental explorations and potential applications.

Silicon telluride ($Si_2Te_3$) is a unique layered semiconductor that has recently been made into a few atomic layer thickness.[17] In $Si_2Te_3$, the Te atoms form a hexagonal close-packed structure, while Si atoms form Si-Si dimers and fill 2/3 of the allowed "metal" sites (Fig. 1a).[18,19] For each Si-Si dimer, there are four possible orientations: 3 in-plane (horizontal) directions and one out-of-plane direction. A recently theoretical studys show that the rotation of Si dimers has an energy barrier of about 1 eV and can happen at room temperature, giving rise to peculiar structure variability.[20] Because the Si-Si dimers rotation is coupled to the change of lattice size, $Si_2Te_3$ is expected to have unusual mechanical properties and other properties such as electronic structure are expected to strongly coupled to mechanical strain as well.

---

[*] Email: xshen1@memphis.edu

In this paper, we report a computation study of the ideal strength of monolayer silicon telluride ($Si_2Te_3$) using first-principles density functional theory (DFT) method. We find that the $Si_2Te_3$ monolayer can sustain a uniaxial tensile strain up to 38%, highest among all two-dimensional materials reported to date. The effect of uniaxial strain over the electronic band structure are investigated. The large critical strain makes it possible to mechanically reduce the band gap by up to 1.4 eV. The materials undergo an unusual indirect-direct-indirect-direct transition with increasing strain is discovered. We have also found that mechanical strain can completely suppress the flipping of Si-Si dimers.

First-principles density functional theory (DFT) calculations were performed using VASP (Vienna Ab initio Simulation Package).[21] The pseudopotential used in the calculations was constructed under the projected augmented wave (PAW) method.[22] We use the Perdew-Burke-Ernzerhof (PBE) exchange-correlation functional[23] under generalized gradient approximation (GGA). The convergences are achieved when the difference in energies in two successive steps becomes less than $10^{-4}$ eV in electronic relaxation and less than $10^{-3}$ eV in atomic relaxation. The kinetic energy cutoff for the plane wave basis set was 245 eV. The sampling of the Brillouin zone was done on a 4×4×1 K-point grids centered at the Γ point. The band structure is investigated by both the standard DFT-PBE method and the hybrid functional Heyd-Scuseria-Ernzerhof (HSE) 06 method.[24, 25] To model the $Si_2Te_3$ monolayer under the periodic boundary condition, a vacuum of 8 Å is inserted between the periodic replicas of the monolayer. After obtaining the relaxed structure of monolayer $Si_2Te_3$, a series of incremental tensile strain up to 40% was applied along y (along the Si-Si dimers) direction, and the lattice constant in the x-direction is fully relaxed. The stress for the monolayer is obtained by rescaling the value obtained from three-dimensional modeling unit cell by a factor $Z/d_0$, where $Z$ is the length of the cell in the z-direction and $d_0$ is the effective thickness of the monolayer, whose values in our calculations are 15.16 Å and 6.97 Å, respectively.

Optimized structure of $Si_2Te_3$ is shown in Figure 1a. The structure corresponds to the ground state where all Si-Si dimers are aligned. The primitive cell is marked by the dotted red lines and contains 4 Si and 6 Te atoms. To facilitate the applications of strain in y-direction and relax the lattice in the perpendicular direction, we use a rectangular cell for calculation as marked by the black lines, which contains 8 Si and 12 Te atoms. To estimate the elastic limit of monolayer $Si_2Te_3$, the stress as a function of tensile strain at y-direction is shown in Figure 1b. This curve tells us that the monolayer $Si_2Te_3$ can sustain a critical tensile strain of 38%. This critical tensile strain is highest among all 2D materials which have been reported.[26-38] Previously, the reported highest critical strain is 30% for boron nitride[34] and phosphorene.[36] Meanwhile, even at such high critical strain, the breaking strength (stress) is 8.63 N/m, which is low among 2D materials. For comparison, the breaking strength for $MoS_2$ is 15 N/m.[39] These results suggest that $Si_2Te_3$ is an extremely flexible 2D material.

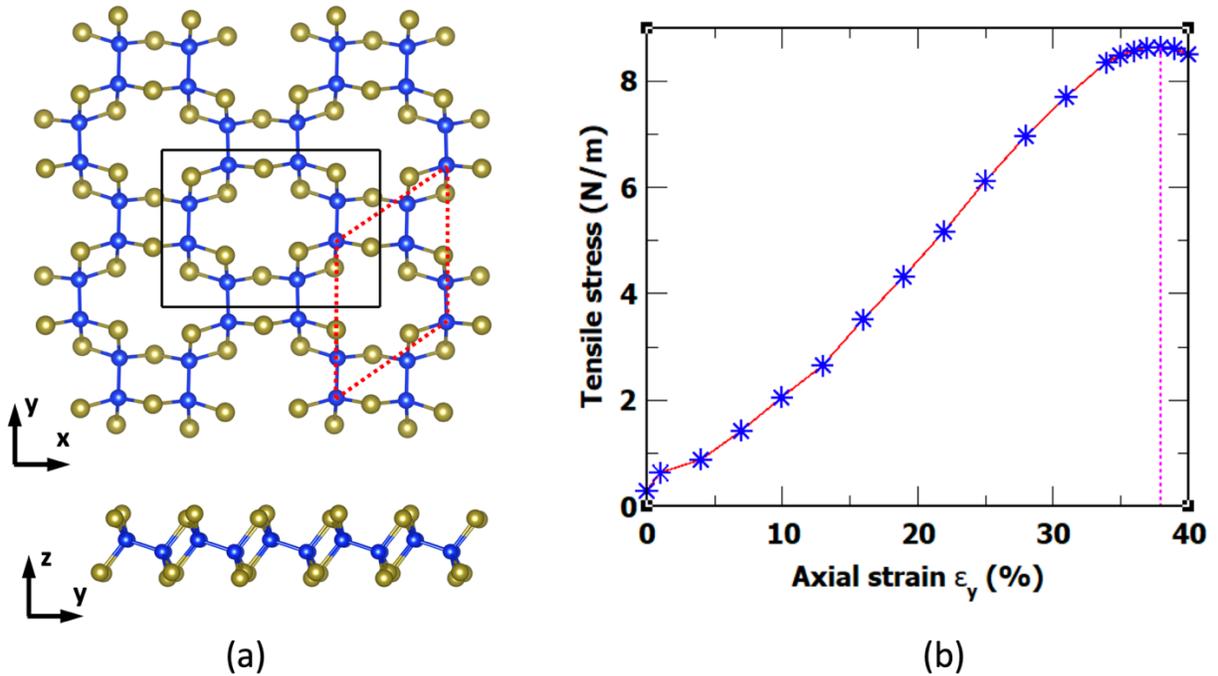

Figure 1. (a) Structure of monolayer $Si_2Te_3$. Si and Te atoms are shown in blue and tan colors respectively. (b) The stress as a function of tensile strain along y axis (the direction of Si-Si) dimers.

The high flexibility enables a large mechanical strain being applied to $Si_2Te_3$ to tune the electronic structure. In Figure 2 we show the band gap as a function of strain using both DFT and hybrid DFT (HSE06) methods. At +34% strain, the band gap changes from 2.6 eV to 1.2 eV in HSE calculations, which is a reduction of 1.4 eV. The DFT calculations yield a similar value of band gap reduction of 1.2 eV. By using the HSE band gap as the baseline and taking into account that 2D semiconductor monolayers can have large exciton binding energy up to 0.6 to 1 eV,[40, 41] it is likely that mechanical strain can tune the exciton emission in $Si_2Te_3$ monolayer from visible to the longer wavelength part of the near-infrared spectrum or even the mid-infrared.

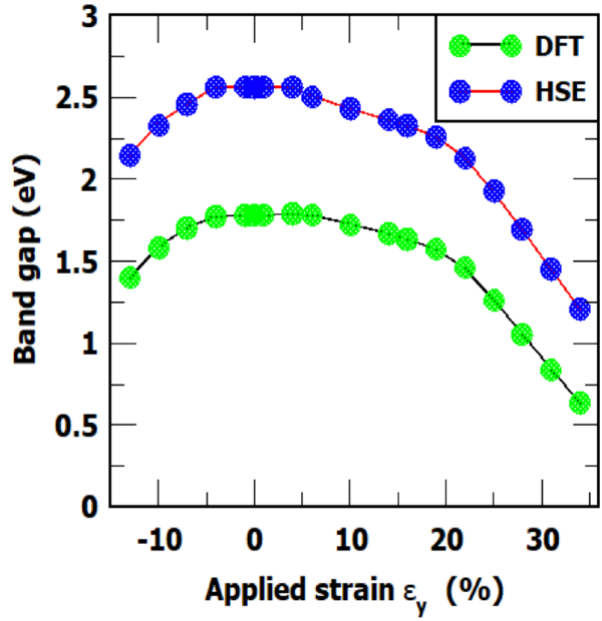

Figure 2. Band gap as a function of applied uniaxial strain for monolayer Si$_2$Te$_3$ using DFT and hybrid DFT (HSE) methods.

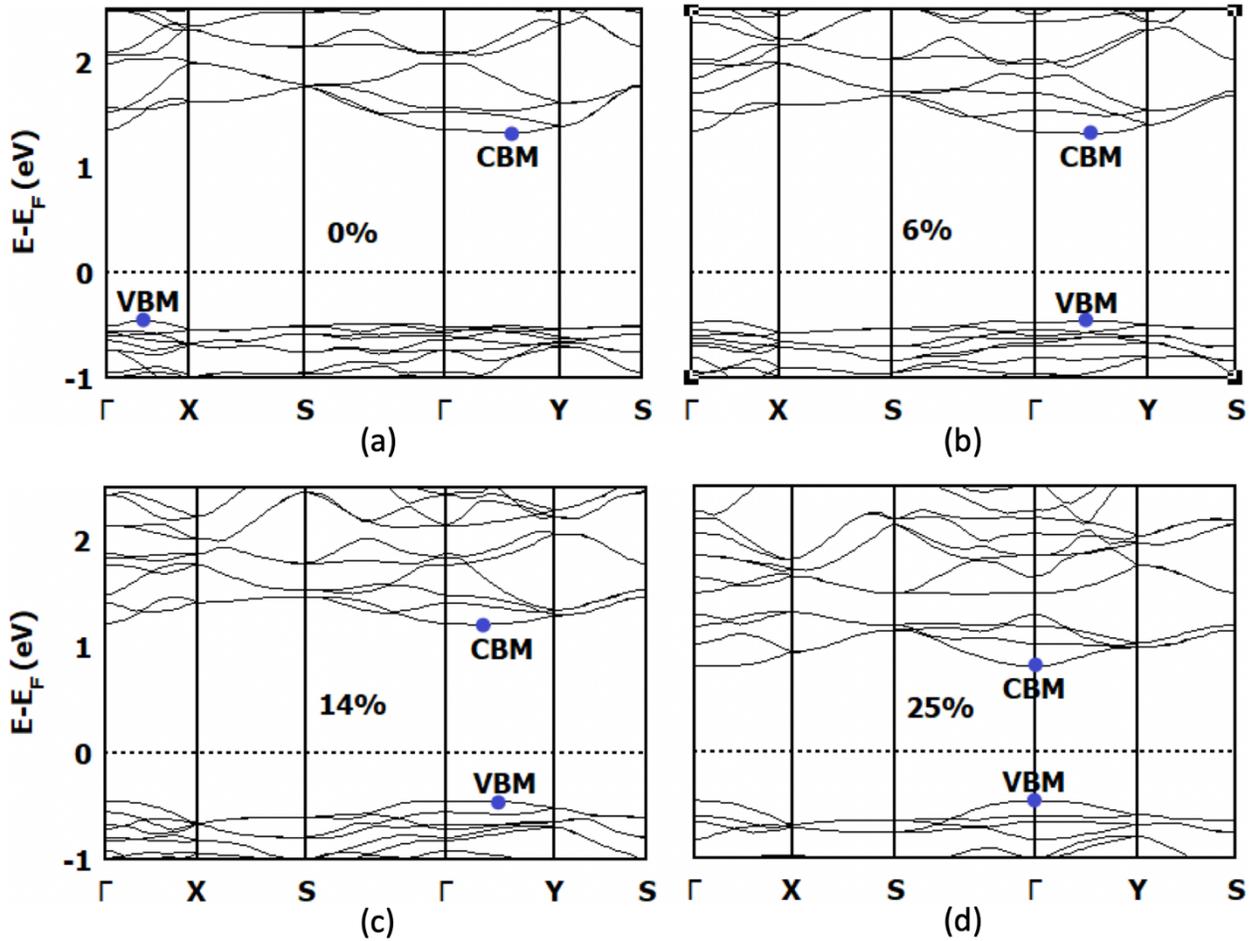

Figure 3. The band structure of monolayer Si$_2$Te$_3$ under uniaxial strain along y-axis.

The mechanical strain not only changes the band gap but also the nature of the band edge transition. Figure 3 shows the band structure of the monolayer $Si_2Te_3$ for different strain along the y-direction calculated using the DFT method. For unstrained structure (0%, Figure 3a), the valence band maximum (VBM) lies between the Γ and X points and the conduction band maximum (CBM) lies between the Γ and Y points, giving rise to the indirect band gap. At +6% strain (Figure 3b), the VBM shifts to a point between Γ and Y and coincides with the CBM, whose position has shifted slightly towards Γ. With increasing strain, the CBM moves further towards Γ point and the band gap becomes indirect again. As shown in Figure 3c, for the case of +14% strain both the VBM and CBM lie between Γ to Y, however, they are not at the same point. Increasing the strain further, both the CBM and VBM shift to the Γ point at 25% strain (Figure 3d), making the band gap direct again. For strain greater than 25%, the CBM and VBM stay at the Γ point band gap remains direct. To the best of our knowledge, this type of indirect-direct-indirect-direct transition in response to mechanical strain hasn't been reported for other 2D materials. Most transition metal dichalcogenide monolayers feature a direct-indirect transitions[42] and 2D phosphorene has a direct-indirect-direct transition.[36] The direct-indirect-direct transition in phosphorene was explained through the competition of the near band edges states, aka, different bands taking the band edge positions at the different strain. For $Si_2Te_3$, it is clear from Figure 3 that the CBM and VBM at each strain level are at the same bands and there is no competition among the nearby bands for the band edge. Thus, such of indirect-direct-indirect-direct transition is likely caused by the modification of chemical bonding in the relatively complex crystal structure.

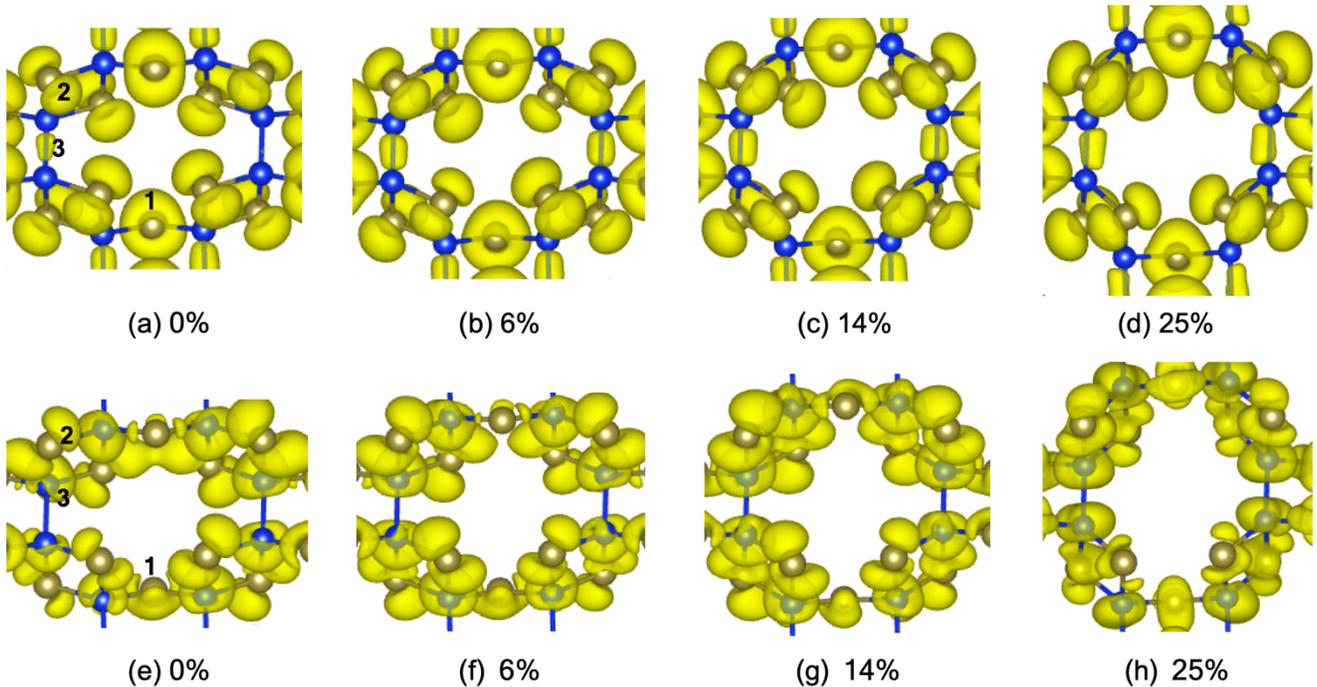

(a) 0%   (b) 6%   (c) 14%   (d) 25%

(e) 0%   (f) 6%   (g) 14%   (h) 25%

Figure 4. The square modulus of the wavefunctions at VBM (a-d) and CBM (e-h) for various strain levels.

To better understand the effect of strain at the electron level, we analyze the electron wave functions of the VBM and CBM for 0%, 6%, 14%, and 25% strain. The VBM (Figure 4a-4d) mainly consists of the non-bonding states of Te 5s orbitals as marked by point 1 and the Te 5p orbitals as marked by point 2, along with a small contribution from the Si-Si bond as marked by point 3. The Te 5s orbitals (point 1) remain unaffected even at sufficiently large strain. The Te 5p orbital (point 2) shows significant change upon the increasing strain. At 0%, the electron wavefunction at point 2 has the character of $p_x + p_y$ (Figure 4a). With increasing strain, the orbital rotates counter-clockwise and becomes more $p_x$-like (Figure 4d). Charge distribution in Si-Si bond at point 3 also exhibits visible changes with the strain. There is little charge density at point 3 at 0% (Figure 4a), suggesting that the Si-Si bond orbital lies below the VBM, which composes mostly of Te 5s and 5p orbitals. As the strain increases to 6% (Figure 4b), the contribution from Si-Si bond to the wavefunction at VBM increases. This indicates that the strain along y-direction is weakening the Si-Si bond and thus raising the energy of this bonding orbital. As strain further increases from 6% to 25%, the contribution from Si-Si bond changes from being symmetric around the Si-Si bond (Figure 4b) to be more profound towards the center of the ring structure (Figure 4d), indicating that Si-Si bond is bearing the load induced by the strain. The existence of the Si-Si bond at the "metal" site is a unique feature of $Si_2Te_3$ and the fact that this additional Si-Si bond is bearing the load may explain why this material can sustain an unusually large tensile deformation of 38%.

The CBM wavefunction as a function of strain is shown in Figure 4e-4h. As shown in Figure 4e, the CBM mainly consists of the non-bonding states of Te 5s orbitals as marked by point 1 and the Te 5p orbitals as marked by point 2, and Si 2s orbitals as marked by point 3. As the strain increase, the Te 5s orbital at point 1 contributes less to the CBM (Figure 4g, 4h), suggesting other orbitals are perturbed more by the strain. Meanwhile, the Te 5p orbitals at point 2 undergoes significant change and becomes more s-like (Figure 4h). The Si 2s orbital at point 3 also undergoes visible changes. At 0% strain, this state is partially excluded in the region of the Si-Si bond (Figure 4e). With increasing strain, it expands to the region of Si-Si bond (Figure 4h), which is consistent with the Si-Si bonding orbital being weakened under tensile strain. As the Si 2s orbital expands towards the Si-Si bond, the gap between the upper and lower half of the simulation cell narrows, resulting in more overlap of wavefunctions in the y-direction along the Si-Si bonds, as can be seen by contrasting Figure 4e and Figure 4h. As a result of the better overlap at high strain, the electron state at CBM becomes more extensive along y-direction and thus more dispersive, consistent with a larger curvature at CBM in the band structure as seen in Figure 3d. The larger curvature means smaller effective masses and higher electron mobility. Therefore, a large mechanical strain can potentially significantly enhance the electron transport properties in $Si_2Te_3$.

The structure variability of Si-Si dimers in $Si_2Te_3$ allows many interesting materials properties. Meanwhile, for some applications, a uniform orientation of Si-Si dimers may be beneficial. We expect the uniaxial strain to promote the alignment Si-Si dimers along with the strain direction. To verify this hypothesis, we investigate the effect of strain on the energy cost of having one Si-Si dimer misaligned with other dimers. We consider the modeling cell containing 20 atoms in which, one out of 4 dimers is rotated by 120° in the plane (Figure 5a). Without strain, having such a horizontally misaligned Si-Si dimer will increase the total energy of the cell by 115 meV. Therefore, at room temperature, the probability of a dimer being be misaligned horizontally is about 1 misaligned dimer out of 80 dimers. The calculations show that the energy cost to horizontally flip a dimer increases with positive strain applied along the dimer direction (Figure 5b). At 16% strain, the energy cost increases to 1.54 eV. Such large energy cost means the horizontal misalignment will only happen for about 1 out of $5\times10^{25}$ dimers. This result indicates that the mechanical strain can be used as an effective way to control the dimer orientation in $Si_2Te_3$.

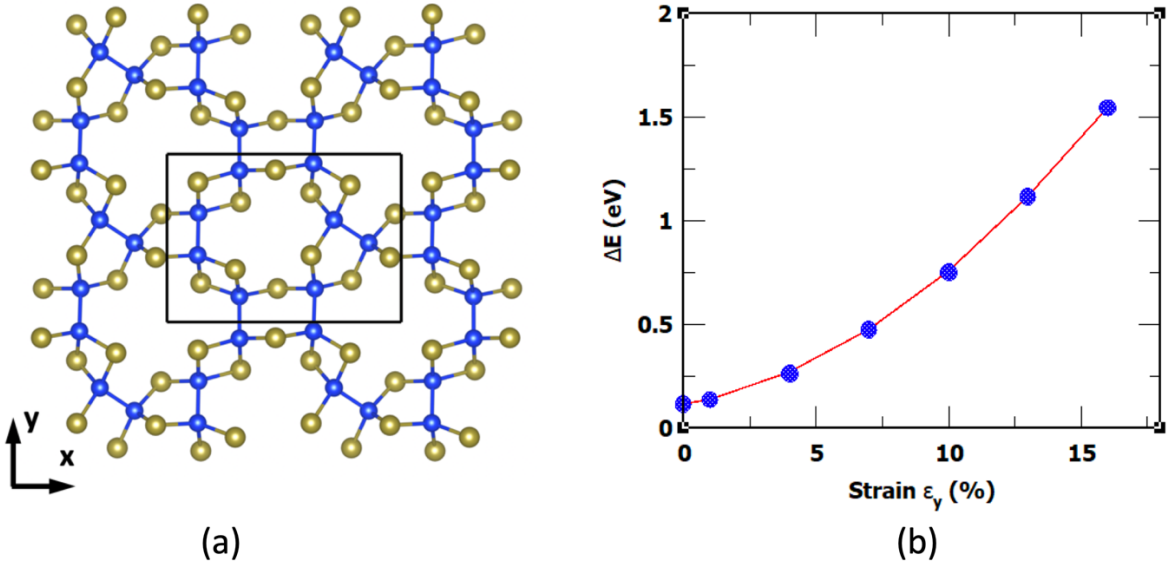

Figure 5. (a) Structure of monolayer $Si_2Te_3$ with one horizontally misaligned dimer (middle) (b) the energy cost to flip a Si-Si dimer as a function of applied strain along y-axis.

In summary, first-principles density functional theory calculations are performed to determine the critical strain of monolayer $Si_2Te_3$. The results show that $Si_2Te_3$ can sustain a critical uniaxial tensile strain up to 38% with a breaking stress of 8.63 N/m along the direction of Si dimer, making $Si_2Te_3$ the most flexible 2D material reported. Because of the high flexibility, a large strain can be used to tune the band structure and the band gap can be reduced by up to 1.4 eV. This tunability will be useful for applications in photonics and optoelectronics from visible to infrared range. With increasing strain, the band gap undergoes an unusual indirect-direct-indirect-direct transition. We also find that the uniaxial strain can effectively control the orientation of Si dimers.

**ACKNOWLEDGEMENTS**

This work was supported by National Science Foundation grant # DMR 1709528 and by the Ralph E. Powe Jr. Faculty Enhancement Awards from Oak Ridge Associated Universities (ORAU). Computational recourses were provided by University of Memphis High-Performance Computing Center (HPCC) and by the NSF XSEDE under grants # TG-DMR 170064 and 170076. We thank Dr. Jingbiao Cui and Dr. Thang B. Hoang for helpful discussion.